# Role of Ca off-centering in tuning the ferroelectric phase transitions in Ba(Zr,Ti)O$_3$ system

Desheng Fu[1,2] and Mitsuru Itoh[3]

[1]Department of Electronics and Materials Science, Graduate School of Engineering, Shizuoka University, 3-5-1Johoku, Hamamatsu 432-8561, Japan. [2]Department of Optoelectronics and Nanostructure Science, Graduate School of Science and Technology, 3-5-1 Johoku, Naka-ku, Hamamatsu 432-8011, Japan. [3]Materials and Structures Laboratory, Tokyo Institute of Technology, 4259 Nagatsuta, Yokohama 226-8503, Japan.

**Abstract:** We here report the substitution effects of the smaller Ca for the bulky Ba in (Ba$_{1-x}$Ca$_x$)(Ti$_{-1-y}$Zr$_y$)O$_3$ perovskite oxides for two systems (Ba$_{1-x}$Ca$_x$)TiO$_3$ with $y$=0 and (Ba$_{1-x}$Ca$_x$)(Ti$_{0.9}$Zr$_{0.1}$)O$_3$ with $y$=0.1. Ca off-centering was found to play a critical role in stabilizing the ferroelectric phase and tuning the polarization states in both systems. It was demonstrated that the atomic displacement due to Ca off-centering in the bulky Ba-site in the perovskite structure provides an effective approach to compensate the reduction of ferroelectricity due to the chemical pressure, which allows to keep the Curie point nearly constant in the (Ba$_{1-x}$Ca$_x$)TiO$_3$ system and increase the Curie point in the (Ba$_{1-x}$Ca$_x$)(Ti$_{0.9}$Zr$_{0.1}$)O$_3$ system. It was commonly observed that the Ca off-centering effects lead to the shift of the *R–O* and *O–T* phase transitions toward lower temperatures and the ferroelectric stability of the *T*-phase, resulting in the occurrence of quantum phase transitions with interesting physics phenomena at low temperatures in the (Ba$_{1-x}$Ca$_x$)TiO$_3$ system and the great enhancement of electromechanical coupling effects around room temperature in the (Ba$_{1-x}$Ca$_x$)(Ti$_{0.9}$Zr$_{0.1}$)O$_3$ system over a wide composition range of the Ca-concentration. These finding may be of great interest for the design of the green piezoelectric materials.

# 1. Introduction

There is increased interest in developing the green-piezoelectric materials in the field of electronics materials due to the environment concern on the toxicity of commercially used lead-based Pb(Zr,Ti)O$_3$ (PZT) piezoelectric ceramics. As listed in table 1(Ref. 1-4), BaTiO$_3$ single crystal shows the highest value of piezoelectric coefficient among the single crystals of lead-free piezoelectric. Although the reported values of its piezoelectric coefficient are rather different, recent investigations on the mono-domain of single crystal by the high energy synchrotron x-ray radiation show that BaTiO$_3$ has a $d_{33}$ value of 149±54 pm/V at least at the level of lattice distortion (Fig.1).[4] The large piezoelectric response makes BaTiO$_3$ as a promising material to develop the novel green piezoelectric ceramics.[5-9]

| Crystal | Point group | $d_{33}$ (pC/N) | $\varepsilon_{33}$ | $k$ (%) |
|---|---|---|---|---|
| quartz | 32 | 2.3($d_{11}$) | 4.6 | 10 (X-cut) |
| ZnO | 6mm | 10.6 | 11 | 41 ($k_{33}$) |
| LiNbO$_3$ | 3m | 6 | 30 | 17 (Z-cut) |
| PbTiO$_3$ | 4mm | 19.3 | 121 | 64($k_{33}$) |
| BaTiO$_3$ | 4mm | 149 | 168 | 65($k_{33}$) |
| PZN-PT8% | 3m[111] | 84 | 1000 | 0.39($k_{33}$) |
| | [001] | 2500 | 5000 | 0.94($k_{33}$) |

**Table 1.** Typical piezoelectric crystals and their piezoelectric ($d_{33}$ or $d_{11}$) & dielectric ($\varepsilon_{33}$) constants, and electromechanical coupling factor $k$.[1-4]

Piezoelectricity is the ability of single crystal with noncentrosymmetry (with the exception of point group 432) to develop an electric charge proportional to a mechanical stress or to produce a deformation proportional to an electric field. The piezoelectricity in BaTiO$_3$ is a direct result of its ferroelectricity that has an origin from the Ti atomic displacement in the oxygen octahedron of the ABO$_3$ perovskite structure[10,11](Fig. 2). As shown in variation of the dielectric permittivity with temperature in Fig. 2, there are two challenging issues remain to be solved for BaTiO$_3$: (1) the temperature instability of physical properties around room temperature due to the tetragonal(*T*)-orthorhombic(*O*) phase transition; (2) its relatively lower Curie point of ~400 K in comparison with the lead-based piezoelectrics. A-site substitution of Pb for Ba is able to increases the Curie point; however, such approach is undesirable for the green piezoelectric. Principally, A-site and/or B-site substitution can be used to modify the ferroelectricity of BaTiO3. Here, we show that A-site substitution of Ca for Ba in the Ba-based perovskite oxides can lead to variety of interesting phenomena: (1) the dramatic improvement of temperature stability of its physical properties, (2) the occurrence of quantum fluctuation in low temperatures, and (3) the great enhancement of the electromechanical responses.[6-9]

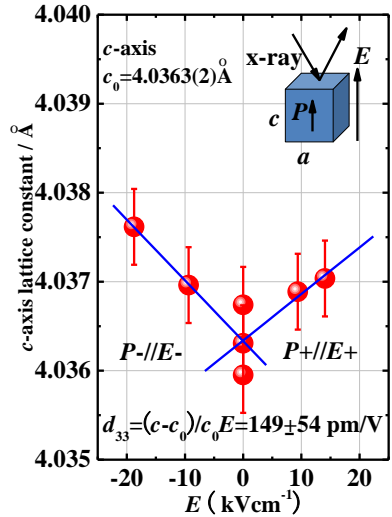

**Figure 1.** Lattice distortion of mono-domain of the BaTiO3 crystal under the application of an electric field.[4]

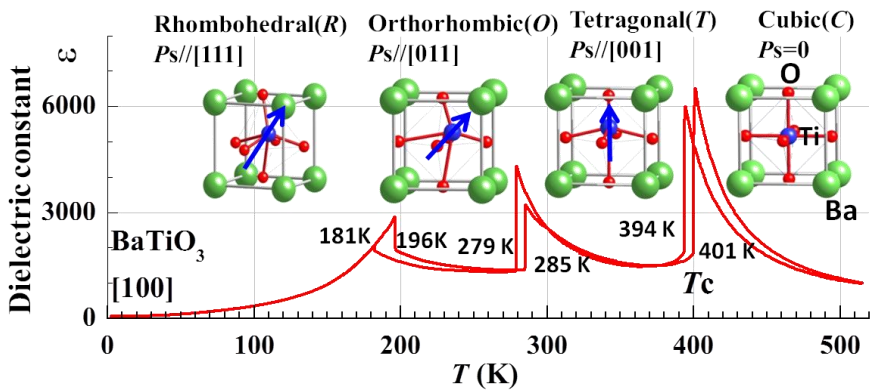

**Figure 2.** Change of dielectric permittivity of the BaTiO3 single crystal with temperature. The schematics of Ti-displacement in the oxygen octahedron of the perovskite structure are also shown.

## 2. Effects of Ca-substitution in BaTiO$_3$

In the early years of 1960, Mitsui and Westphal investigated the influence of Ca-substitution on the phase transition ins BaTiO$_3$.[12] Using the ceramic samples, they have established a phase diagram of (Ba$_{1-x}$Ca$_x$)TiO$_3$ in the composition range of x<0.25 mole for temperatures higher than 73 K. One interesting finding is that the Curie point remains nearly unchanged within the studied composition range. Such behavior is unexpected when considering that CaTiO$_3$ is paraelectric and ionic radius of Ca (~1.34Å) is smaller than that of Ba (~1.60Å),[13] which will lead to the shrink of the unit cell and is expected to reduce the ferroelectricity of the system as substituting Ca for Ba. To gain a new insight into the role of Ca-substitution in the Ba-based perovskite oxides, we have re-examined (Ba$_{1-x}$Ca$_x$)TiO$_3$ system by using single crystal samples, which allow us to observe the intrinsic phenomena of the system.[6-8,14]

### 2.1. Crystal growth

To obtain the single crystal of (Ba$_{1-x}$Ca$_x$)TiO$_3$, we used the floating zone (FZ) technique (Fig. 3(a)) that allows us to grow the single crystal with high purity. According to the reported phase equilibria in the system (1-$x$)BaTiO$_3$-$x$CaTiO$_3$ (Fig. 4), only the crystal with a congruently melting composition ($x$=0.27 report by DeVries and Roy[15] or $x$=0.227 report by Kuper et al[16]) can be directly grown from the melt. Surprisingly, we can grow single crystal of (Ba$_{1-x}$Ca$_x$)TiO$_3$ with perovskite structure for a wide composition range of 0.02≤$x$≤0.34 with a high growth rate of 20 mm/h by the FZ technique.[6] Fig. 3 (b) shows a rod of crystal obtained by this method. It was found that the crystal can be stably grown under the atmosphere of Ar or N$_2$ gas. The crystal is yellowish but transparent. Laue X-ray diffraction patterns clearly indicate that the obtained (Ba$_{1-x}$Ca$_x$)TiO$_3$ crystal has a perovskite structure.

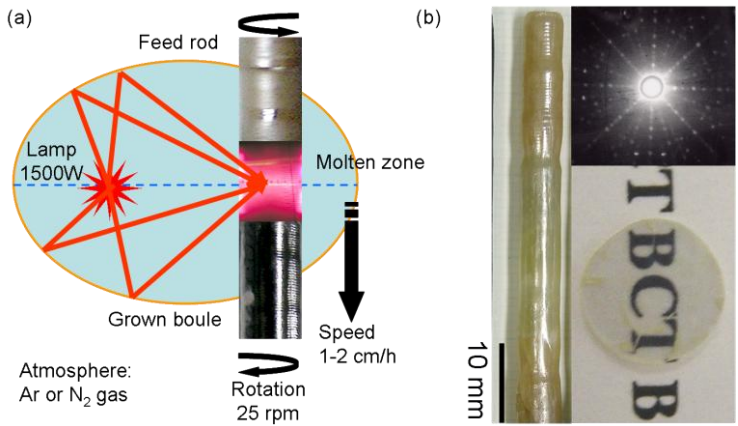

**Figure 3.** (a) A schematic drawing of a floating zone (FZ) technique to grow the (Ba$_{1-x}$Ca$_x$)TiO$_3$ single crystal. (b) Photograph of a (Ba$_{1-x}$Ca$_x$)TiO$_3$ single crystal grown by the FZ technique and its Laue X-ray back diffraction patterns along [001]$_c$ direction of perovskite structure.

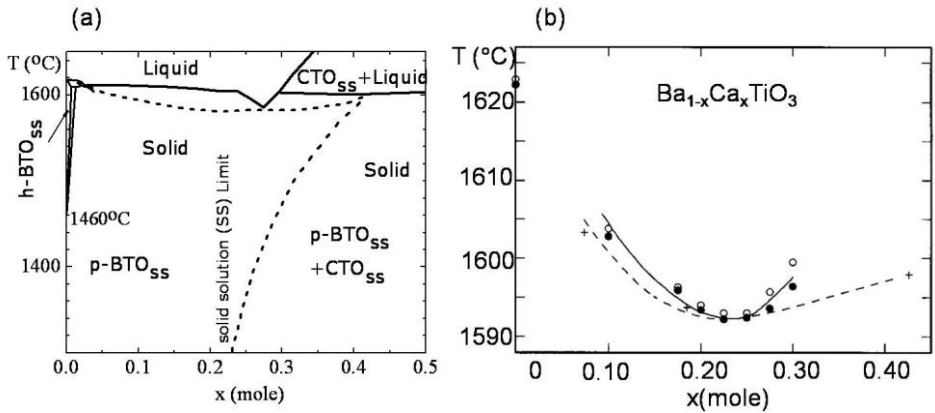

**Figure 4.** Phase equilibria in the system $(1-x)$BaTiO$_3$-$x$CaTiO$_3$ reported by (a) DeVries and Roy[15] and (b) Kuper et al[16].

## 2.2. Dielectric behaviours

Figure 5 shows the temperature dependence of dielectric permittivity of the (Ba$_{1-x}$Ca$_x$)TiO$_3$ single crystals in a temperature range from 2K to 400K. Compared with the polycrystalline ceramics, the (Ba$_{1-x}$Ca$_x$)TiO$_3$ single crystal show very sharp change of dielectric response at the phase transition, allowing to easily determine the transition temperatures. Similar to the polycrystalline ceramics[12], the Curie point is nearly independent on the Ca-concentration. However, the tetragonal (*T*)-orthorhombic (*O*) and *O*-rhombohedral (*R*) ferroelectric transitions are shifted to lower temperatures as increasing the Ca-concentration. For compositions of $x>0.23$, these two transitions completely disappear, and *T*-phase is the only stable ferroelectric phase in the crystal. This situation is very similar to that of PbTiO$_3$, in which the *T*-phase is the only stable ferroelectric structure. Another interesting thing is that the dielectric permittivity is nearly unchanged for the temperatures lower than the Curie point for $x>0.23$. This unique behavior provides a possibility to design electronic devices operating stably in a wide temperature range from boiling point of water to absolute zero Kelvin by using (Ba$_{1-x}$Ca$_x$)TiO$_3$.

As well-known, the dielectric response in the displacive-type ferroelectric is dominated by the phonon dynamics, particularly the soft-mode behavior. Lyddane-Sachs-Teller (LST) relationship predicts that the dielectric permittivity is inversely proportional to the soft-mode frequency. To understand the temperature independent of dielectric response of (Ba$_{1-x}$Ca$_x$)TiO$_3$ ($x >0.23$). We have performed confocal micro-Raman scattering measurements for the (Ba$_{1-x}$Ca$_x$)TiO$_3$ ($x =0.233$) single crystal to clarify its soft mode dynamics.[14] In contrast to BaTiO$_3$, a well-defined soft phonon mode was observed for temperatures lower than Curie point in the (Ba$_{1-x}$Ca$_x$)TiO$_3$ ($x =0.233$) single crystal. The temperature dependence of the soft-mode frequency agrees qualitatively with the dielectric permittivity through Lyddane-

Sachs-Teller relationship (Fig.6). This result clearly indicates that the unique dielectric response of $(Ba_{1-x}Ca_x)TiO_3$ ($x$=0.233) is directly derived from its soft-mode dynamics.

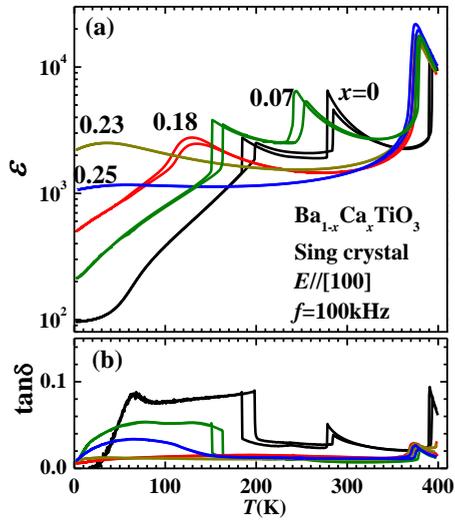

**Figure 5.** Dielectric behaviours of the $(Ba_{1-x}Ca_x)TiO_3$ single crystals.

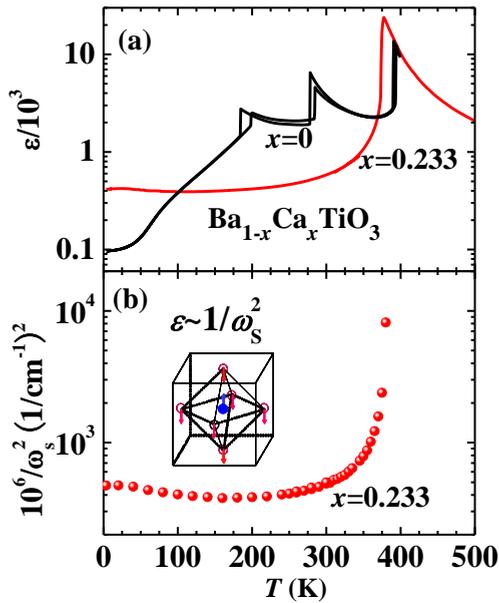

**Figure 6.** Comparison of (a) dielectric permittivity of the $(Ba_{1-x}Ca_x)TiO_3$ single crystal and (b) the phonon frequency of Slater soft-mode.

## 2.3. Phase diagram and quantum phase transitions

From the temperature change of dielectric permittivity of the $(Ba_{1-x}Ca_x)TiO_3$ crystal, we have established its phase diagram in a composition range of $x \leq 0.34$ for temperature down to 2 K (Fig. 7). Compared with the phase diagram proposed by Mitsui and Westphal using ceramics,[12] our phase diagram has been expanded to a composition up to $x=0.34$ mole and a low temperature down to 2 K. These expansions of composition and temperature allow us to reveal some unexpected phenomena in this system: (1) ferroelectric $R$- and $O$-phases become unstable as increasing the Ca-concentration, and are predicted to disappear for $x > x_c^{O-R} = 0.18$ and $x > x_c^{T-O} = 0.233$, respectively; (2) ferroelectric $T$-phase is a ground state for $x > x_c^{T-O}$.

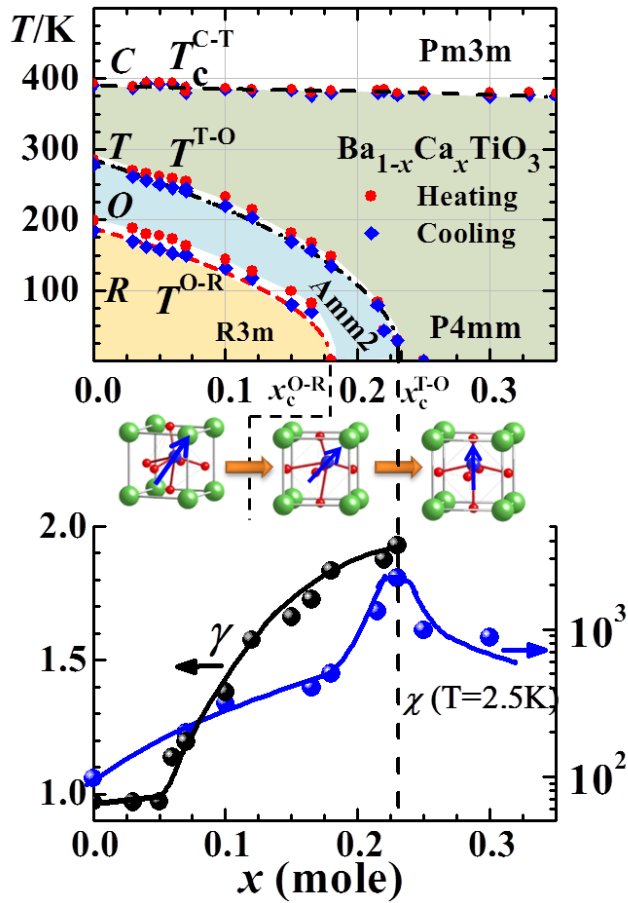

**Figure 7.** Top panel: phase diagram of the $(Ba_{1-x}Ca_x)TiO_3$ crystals. Left of bottom panel: Change of critical exponent $\gamma$ for the dielectric susceptibility in the $T$-$O$ phase transition. Right of bottom panel: Compositional dependence of the dielectric susceptibility measured at 2.5 K.

One important finding is that ferroelectric-ferroelectric quantum phase transitions occur in the $(Ba_{1-x}Ca_x)TiO_3$ crystals. The occurrence of ferroelectric-ferroelectric quantum phase transition is supported by two experimental facts: the compositional dependence of $T^{O-R}$ and $T^{T-O}$ transition temperatures and the temperature dependence of the dielectric susceptibility in the crystals with composition close to $x_c^{T-O}$. Theoretical and experimental studies[17-19] on quantum phase transition indicate that (a) for a quantum ferroelectric, the transition temperature depends on the substitution concentration (i. e., on an effective order parameter) as

$$T_c \propto (x-x_c)^{1/2} \quad (1),$$

as opposed to the classical relationship,

$$T_c \propto (x-x_c) \quad (2).$$

(b) The inverse dielectric susceptibility varies with temperature as

$$\chi^{-1} \propto (T-T_c)^2 \quad (3)$$

for the quantum mechanical limit instead of the classical Curie-law

$$\chi^{-1} \propto (T-T_c) \quad (4).$$

Our phase diagram clearly shows that the *T-O* and *O-R* phase transitions deviate from the classical relationship (equation (2)) for $x<0.06$, and exactly follow equation (1) for $x>0.06$ with $x_c$ equal to $x_c^{T-O}=0.233$ and $x_c^{O-R}=0.18$, respectively.

To examine point (b), we have analyzed the temperature variation of the inverse dielectric susceptibility of *T-O* phase transition in the $(Ba_{1-x}Ca_x)TiO_3$ crystals with the following equation,

$$\chi^{-1} \propto (T-T_c)^\gamma \quad (5).$$

Figure 8 shows two typical examples: one for $x=0$, and another for $x=0.23$ close to $x_c^{T-O}$. For pure $BaTiO_3$ with $x=0$, the classical Curie law with $\gamma=1$ was observed. In contrast, for $x=0.23$, the critical exponent $\gamma$ for the susceptibility was found to have a value of 2, which is predicted for the quantum phase transition (equation (3)). The left of bottom panel in Fig. 7 shows the variation of $\gamma$ with $x$. It is clear that the $\gamma$ changes from the value of classical limit to that of quantum limit as increasing $x$ from 0 to $x_c^{T-O}$. This fact again indicates that a quantum phase transition indeed occur at zero Kelvin in the system when changing the Ca-concentration. An interesting phenomenon is that dielectric anomaly was observed for the *T-O* quantum phase transition at $x=x_c^{T-O}$ as shown in right of bottom panel of Fig 7. At a temperature of 2.5 K close to zero Kelvin, the crystal with $x=0.23$ close to $x_c^{T-O}$ shows a maximum value of dielectric susceptibility in the system.

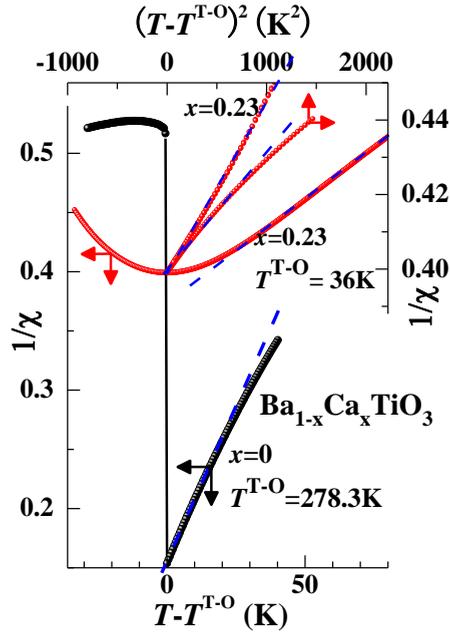

**Figure 8.** Change of the inverse dielectric susceptibility ($\chi=\varepsilon$-1) near the *T-O* phase transition in the (Ba$_{1-x}$Ca$_x$)TiO$_3$ crystals.

## 2.4. Ca off-centering predicted from first principles calculations

As mentioned above, the ionic radius of Ca is approximately 16% smaller than that of Ba. The substitution of Ca for Ba will absolutely results in the shrink of the perovskite unit cell. As shown in Fig. 9(a), both *a*- and *c*-axes of the tetragonal structure shrink as increasing Ca-concentration, resulting in reduction of the unit cell. The volume of unit cell of *x*=0.34 is approximately 3.6% smaller than that of pure BaTiO$_3$. This chemical-pressure-induced reduction of the unit cell doesn't have significant influence on the Curie point ($T_c^{C-T}$) of the system (Fig. 9(c)).

In contrast, for the case of hydrostatic pressure, at the same level of unit-cell reduction, the Curie point is reduced to ~180 K, which is greatly lower than ~400 K of pure BaTiO$_3$ (Fig. 9(c)). The hydrostatic pressure gradually reduces the Curie point, leading to the complete disappearance of ferroelectricity in BaTiO$_3$ at a level of 5% reduction of the unit cell. Although the chemical substitution of the smaller Ca for the bulky Ba also leads to the reduction of the unit cell, the effects of chemical-pressure on the ferroelectricity in the (Ba$_{1-x}$Ca$_x$)TiO$_3$ system are greatly different from what we have expected in the hydrostatic pressure application.

Apparently, the reduction of unit cell by the chemical pressure shrinks the oxygen octahedron in the perovskite structure, resulting in the reduction of available space for Ti off-centering shift in the oxygen octahedron. Since the ferroelectricity is derived from the Ti-shift in the oxygen octahedron in the perovskite structure of $BaTiO_3$, it is naturally expected that chemical-pressure-induced reduction of the unit cell in the $(Ba_{1-x}Ca_x)TiO_3$ system should weaken the ferroelectricity of the system and result in the decrease in its Curie point. However, as shown in the phase diagram in Fig. 7 and Fig. 9, the Curie point remains nearly unchanged with the Ca-substitution. Also, as shown in Fig. 9 (b), the tetragonality ($c/a$), which is generally used to estimate the atomic displacement and thus the ferroelectricity and polarization in the tetragonal ferroelectric,[20,21] keeps a constant value in the whole composition range in the $(Ba_{1-x}Ca_x)TiO_3$ system. In accordance with such unchanged tetragonality, we really have observed that the saturation polarization is also nearly independent on the Ca substitution within the limit of solid solution (see Fig. 11)[8]. These facts suggest that in addition to the Ti-displacement in the oxygen octahedron, there should be an additional atomic displacement to contribute to the ferroelectricity of the $(Ba_{1-x}Ca_x)TiO_3$ system. The ionic radii of Ca and Ba are 1.34 Å and 1.61 Å, respectively. There is a difference of 0.27 Å between them. This suggests that the smaller Ca ion is very possible to have an off-centering displacement in the bulky Ba-site (Fig. 10(b)).

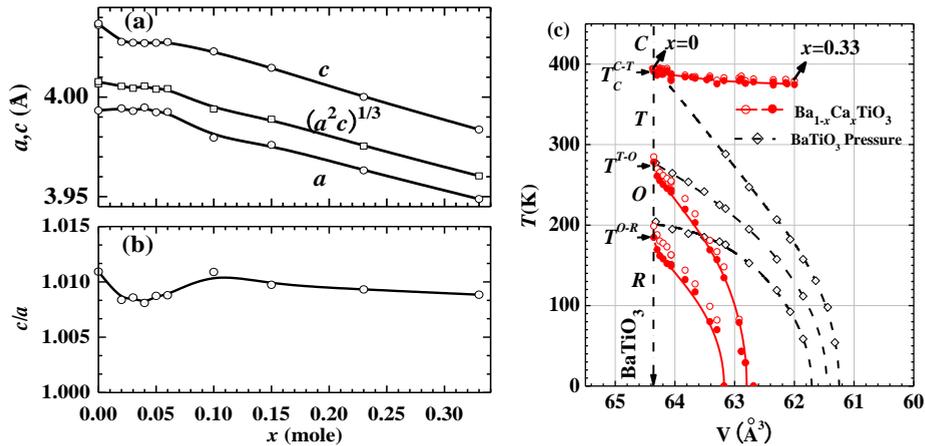

**Figure 9.** Change of (a) lattice constants and (b) tetragonality ($c/a$) of the $(Ba_{1-x}Ca_x)TiO_3$ crystal. (c) Change of phase transition temperature as a function of unit cell volume (determined at room temperature); The results of hydrostatic pressure effects for pure $BaTiO_3$ is also shown for comparison.[7,22,23]

To examine the ideal of Ca off-centering in the bulky Ba-site, we have performed the first principles calculations for this system.[7,24] Since Ca-substitution tends to stabilize the tetragonal structure, we therefore focus on the calculations in this structure to get information about Ca displacement. The results are summarized in Fig. 10. As shown in Fig. 10(a), Ti-shift along the [001] direction leads to the stability of the tetragonal phase in $BaTiO_3$

($x=0$). In our calculations for a substitution of Ca for Ba ($x=1/8$), we have calculated the relative change in potential energy as changing the Ca-location shown schematically in Fig. 10(c). As shown in Fig. 11(d), the Ca off-centering shift results in the lowering of the potential energy of the system. This result clearly indicates that the Ca off-centering stabilizes the structure of the $(Ba_{1-x}Ca_x)TiO_3$ system. After tracing the variation of potential energy for moving Ca along various paths, we found that the mostly possible direction for Ca-shift is [113] since the potential energy is the lowest when Ca is shifted along this direction (Fig. 10(e)). The Ca-shift along the [113] direction seem to be incompatible with the tetragonal structure. However, if we consider the eight-site model similar to that assumed for Ti displacement in $BaTiO_3$, namely, Ca may displace along the equivalent directions [113], [1-13], [-113], [-1-13], or [11-3], [1-1-3], [-11-3], [-1-1-3]. The activation barrier for Ca moving between these equivalent states has been evaluated to be less than 3 meV. Therefore, thermal and spatial averaging among these states allows the preservation of the overall tetragonal symmetry detected from X-ray diffractions. It should be noticed that the estimated displacement of Ca is approximately 0.1 Å (Fig. 10(d)), which is larger than the 0.05 Å shift of Ti in the tetragonal structure of $BaTiO_3$ (see Ref. 1 and Fig. 10(a)).

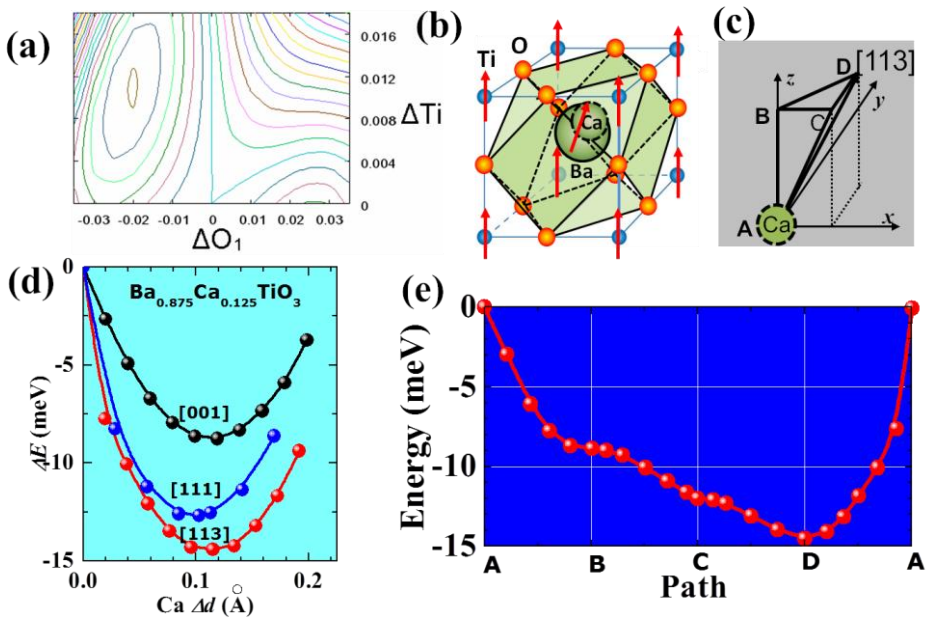

**Figure 10.** (a) Two-dimensional contour map of potential energy of $BaTiO_3$ as function of Ti and $O_1$ displacement along [001] direction of the polar *c*-axis. (b) Schematic of Ca off-centering in the bulky Ba-site of perovskite structure, in which the atomic shifts are shown by the arrows. (c) Direction of Ca-shift for the first principles calculations of $Ba_{7/8}Ca_{1/8}TiO_3$. (d) Change of potential energy of $Ba_{7/8}Ca_{1/8}TiO_3$ along the [001], [111] and [113] directions. (e) Change of potential energy of $Ba_{7/8}Ca_{1/8}TiO_3$ along the path shown in (c).

## 2.5. Polarization and strain responses

For many technical applications, understanding the physical properties of ceramics sample is of great importance. Figure 11 show the variation of polarization, bipolar- and unipolar-field induced strains with the Ca-substitution in the $(Ba_{1-x}Ca_x)TiO_3$ ceramics, which were measured at room temperature. One interesting finding is that the saturation polarization is nearly insensitive to the Ca substitution within the limit of solid solution[8] as shown in Fig. 11(a). This finding is predictable when considering the composition dependence of tetragonality in the $(Ba_{1-x}Ca_x)TiO_3$ system. As shown in Fig. 9(b), the tetragonality has an approximate value of 1.01 within the solid solution limit. As mentioned above, the tetragonality of the ferroelectric perovskite oxides is predicted to be proportional to its spontaneous polarization by the theoretical calculations. [20, 21]

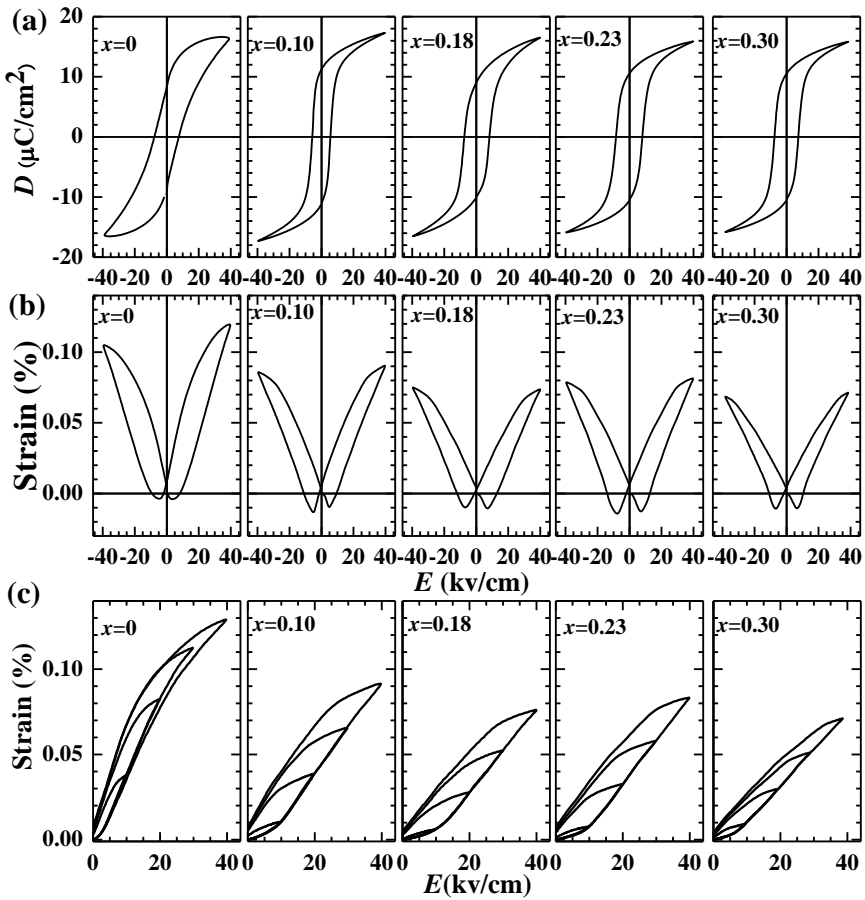

**Figure 11.** (a) Polarization, (b) bipolar-field and (c) unipolar-field strains under electric field in the $(Ba_{1-x}Ca_x)TiO_3$ ceramics.

For the strain response under an electric field, with the exception of BaTiO3 ($x$=0), (Ba$_{1-x}$Ca$_x$)TiO3 ceramics show nearly the same level of strain response under the same electric field. For examples, the strain was observed to be approximately 0.08% for a unipolar field of 40 kV/cm, which corresponds to a level of piezoelectric response of 200 pm/V. The observation of the large strain response in BaTiO3 is not surprised because its *T-O* phase transition is located at a temperature close to room temperature. Around the phase transition, the large response of physical properties generally occurs.

## 3. Effects of Ca-substitution in the Ba(Ti,Zr)O3 solid solution.

To confirm the effects of Ca off-centering in Ba-based perovskite oxides, we also have performed investigations on the system of the Ba(Ti,Zr)O3 solid solution, which have been intensively studied from the middle of 1950s.[25,26] The most amazing findings in this system is that it demonstrates very large piezoelectric response comparable to that of the industrial PZT. High values of electromechanical coupling factor of 74% and large piezoelectric coefficients of 340 pC/N under high field were observed in this system.[27,28] Another interesting thing in this system is that the *O*- or *R*-phase can be tuned to room temperature through controlling Zr-substitution, which is of great significance for the ferroelectric phase modification. However, as shown in the phase diagram reported by Kell and Hellicar (Fig. 12),[25] the problem of this system is that the Curie point decreases with the increase in the Zr-concentration. For Zr-concentrations larger than 20 mol%, the Curie point is reduced to a temperature lower than room temperature, leading to the disappearance of ferroelectricity in the crystal at room temperature. Here, we show that the Ca off-centering effects mentioned above also can be used to increase the Curie point of Ba(Ti,Zr)O$_3$ and enhance its electromechanical coupling effects through tuning the ferroelectric phase boundaries to room temperature.[9]

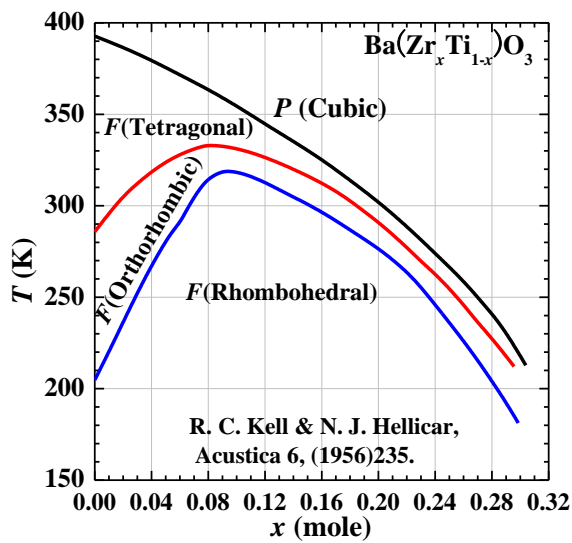

**Figure 12.** Phase diagram of Ba(Ti$_{1-x}$Zr$_x$)O3 solid solutions proposed by Kell and Hellicar.[25]

### 3.1. Sample preparation

In our study, we selected the composition with 10 mol% Zr concentration at which the three successive phases tend to approach each other as shown in Fig. 12. We prepared the (Ba$_{1-x}$Ca$_x$)(Ti$_{0.9}$Zr$_{0.1}$)O$_3$ (BCTZO) ceramics by a solid-state reaction approach. Mixtures of BaCO$_3$, CaCO$_3$, ZrO$_2$ and TiO$_2$ were calcined at 1823 K for 3 h. The calcined powders were ground, pressed and sintered at 1823 K for 5 h. The ceramic pellets were then electroplated with silver for electrical measurements.

### 3.2. Phase formation and structure transformation at room temperature

For the sintering temperature of 1823 K, the single phase of BCTZO was found to be formed within the composition range of $x \leq 0.18$ beyond which non-ferroelectric phase with CaTiO$_3$-type orthorhombic structure occurs and coexist with the BaTiO$_3$-type ferroelectric phase. The phase equilibria of (1-$x$)Ba(Zr$_{0.1}$Ti$_{0.9}$)O$_3$-$x$CaTiO$_3$ is very similar to that of (1-$x$)BaTiO$_3$-$x$CaTiO$_3$ reported by (a) DeVries and Roy[15] as shown in Fig. 4(a). However, the solid solution limit of (1-$x$)Ba(Zr$_{0.1}$Ti$_{0.9}$)O$_3$-$x$CaTiO$_3$ is approximately half of that of (1-$x$)BaTiO$_3$-$x$CaTiO$_3$. This fact indicates that the substitution of Zr for Ti in BaTiO$_3$ will reduce the substitution amount of Ca for Ba.

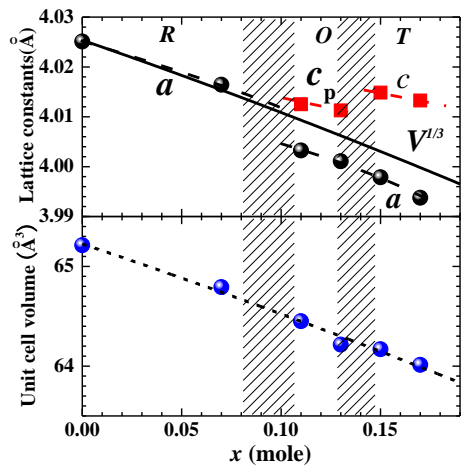

**Figure 13.** Change of the lattice parameters with composition at room temperature for the (Ba$_{1-x}$Ca$_x$)(Ti$_{0.9}$Zr$_{0.1}$)O$_3$ system.

At room temperature, BCTZO with $x=0$ has the ferroelectric rhombohedral structure as shown in phase diagram of Fig. 12. When Ba is substituted with Ca, the structure of BCTZO at room temperature was found to transform from $R$-phase to $O$-phase, and finally to $T$-phase with the increase in Ca-concentration. Figure 13 shows the change of lattice parameters with Ca-concentration for the BCTZO system. Similar to that of $(Ba_{1-x}Ca_x)TO_3$ (Fig. 9(a)), the unit cell of BCTZO shrinks with the substitution of the smaller Ca for the bulky Ba, and its volume is reduced from 65.21 Å$^3$ for $x=0$ to 63.91 Å$^3$ for $x=0.18$. The ferroelectric lattice distortion at room temperature is very small in the BCTZO solid solution. The distortion angles $\alpha$ in the $R$-phase and $\beta$ of the monoclinic unit cell in the $O$-phase have deviations of only 0.01° and 0.1° from the right angle, respectively, while the tetragonality $c/a$ has a value of 1.005 for $x \geq 0.15$ in the $T$-phase.

### 3.3. Phase evolution with temperature

To understand the phase evolution in the BCTZO system, we have measured the temperature variation of dielectric permittivity for different Ca-concentrations. The results are summarized in Fig. 14. As reported in many literatures, due to the approach of the phase transitions, the three successive phase transitions are not easy to be distinguished for BCTZO with $x=0$. However, these $C-T$, $T-O$ and $O-R$ phase transitions as observed in BaTiO$_3$ is clearly demonstrated from the temperature differentiation of dielectric permittivity as shown in Fig. 14. Here, we use the peak of the temperature differentiation of dielectric permittivity to determine the transition temperature of $T-O$ and $O-R$ phase transitions.

The phase diagrams as functions of Ca-concentration and unit cell volume are shown in Fig. 15 (a) and (b), respectively. For comparison, the phase diagram of $(Ba_{1-x}Ca_x)TiO_3$ is also shown in the figure. There are some of similarities between BCTZO and $(Ba_{1-x}Ca_x)TiO_3$ systems: (a) Substitution of the smaller Ca for the bulky Ba shifts the $T-O$ and $O-R$ phase transitions to low temperatures, in other words, Ca-substitution results in the ferroelectric instability of the $O$- and $R$-phases in both systems; (B) In contrast, Ca-substitution enhances the ferroelectric stability of the $T$-phase; (c) The chemical-pressure-induced shrink of the unit cell doesn't reduce the Curie point and weakens the ferroelectricity of both systems. These similarities between BCTZO and $(Ba_{1-x}Ca_x)TiO_3$ systems indicate that Ca off-centering effects play a critical role in tuning the polarization states in these two systems.

However, there are also some of differences between BCTZO and $(Ba_{1-x}Ca_x)TiO_3$ systems: (a) In $(Ba_{1-x}Ca_x)TiO_3$, the Curie point shows a slightly decrease with the increase of Ca-concentration, but it is increased in the BCTZO system. The Curie point is increased from 363 K for $x=0$ to 376 K for $x=0.1$, after which it seems to reach saturation with further substitution in BCZTO; (b) In $(Ba_{1-x}Ca_x)TiO_3$, the $O$- and $R$-phases completely disappear for Ca-substitution amount of $x>0.233$, in contrast, in BCTZO, the disappearance of the $O$- and $R$-phases doesn't occur within the solid solution limit and $R$-phase is still the ground state as occurred in pure BaTiO$_3$. These facts suggest that the contribution of Ca off-centering displacement to the whole spontaneous polarization in BCTZO system is very possible to be greater than that in the $(Ba_{1-x}Ca_x)TiO_3$ system. This interpretation seems to be reasonable. Since $BaZrO_3$ is not ferroelectric even at zero Kelvin and substitution of Zr for Ti reduces the ferroelectricity of $Ba(Ti,Zr)O_3$, in contrast to the large Ti-displacement in the oxygen octahedron, the same level of Zr-displacement is not expected to exist in the oxygen octahedron in $Ba(Ti,Zr)O_3$. Actually, this has been predicted from recent first-principles calculations, which indicates that Zr-displacement is extremely small, and has a value of about sixth of Ti-displacement at the lowest temperature in $Ba(Ti,Zr)O_3$.[29,30] In contrast, as shown in Fig.10. Ca-displacement is predicted to have a value of two times of Ti-displacement from first principles calculations. Therefore, in the BCTZO system, the polarization due to Ca-displacement is able to effectively compensate the reduction of polarization from B-site atomic displacement due to the substitution of Zr for Ti, leading to the enhancement of ferroelectricity in BCTZO system.

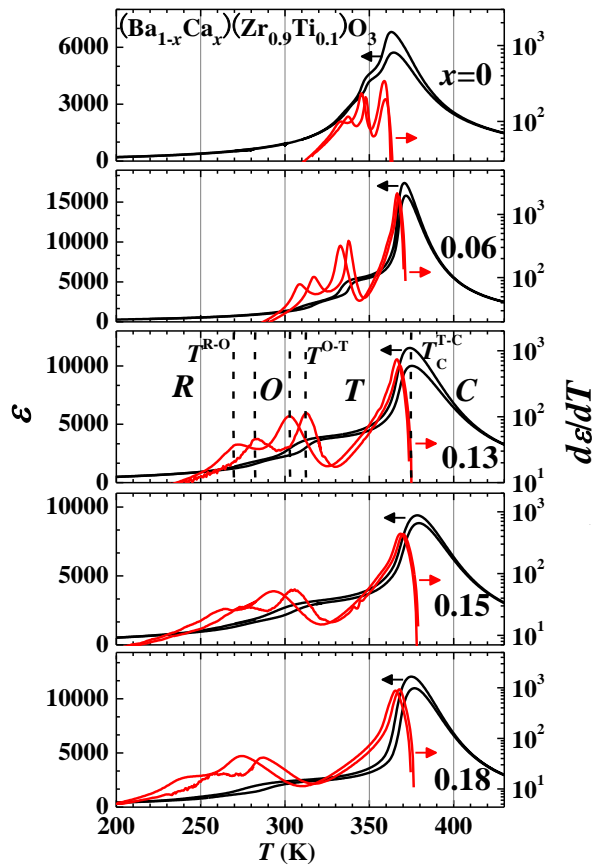

**Figure 14.** Temperature dependence of dielectric permittivity and its temperature differentiation of the $(Ba_{1-x}Ca_x)(Ti_{0.9}Zr_{0.1})O_3$ ceramics.

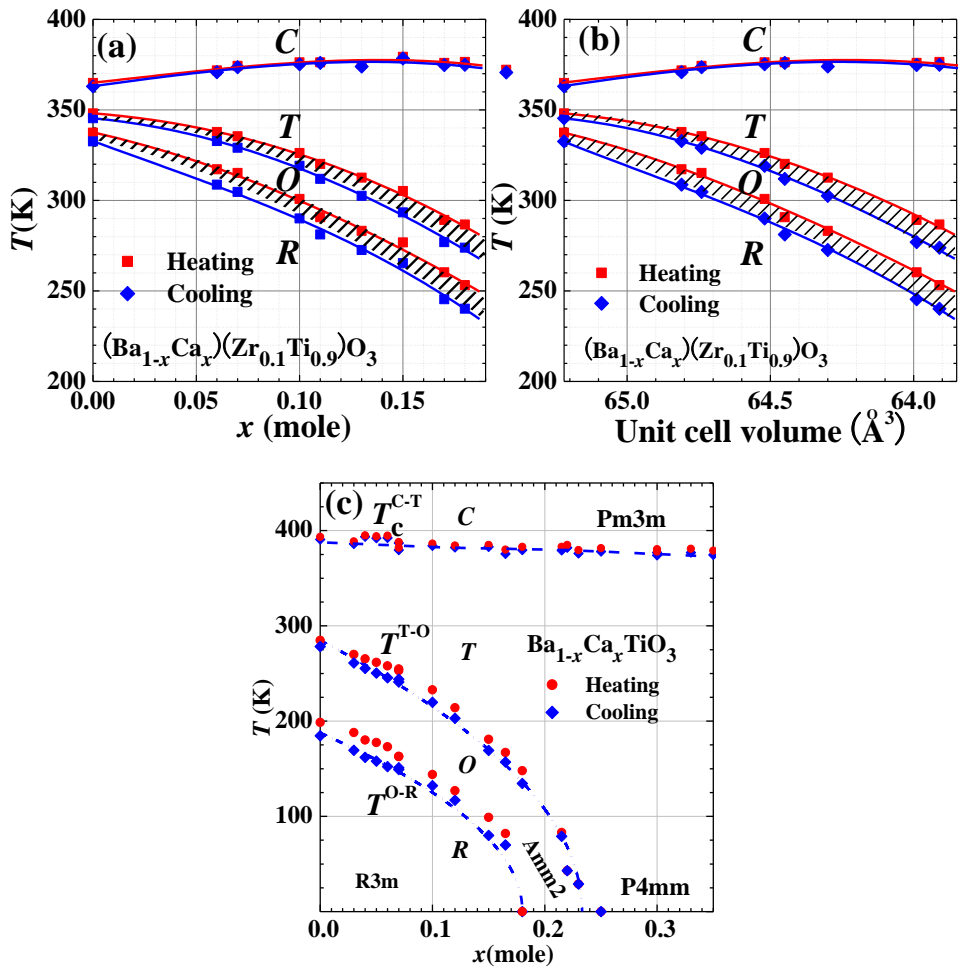

**Figure 15.** Phase diagrams of the (Ba$_{1-x}$Ca$_x$)(Ti$_{0.9}$Zr$_{0.1}$)O$_3$ solid solutions as functions of (a) composition and (b) unit cell volume. For comparison, the phase diagram of (Ba$_{1-x}$Ca$_x$)TiO$_3$ is also shown in (c).

### 3.4. Polarization and strain responses under an electric field

The *D-E* hysteresis loop of BCTZO is shown in Fig. 16 (a). The remanent polarization was observed to have a value of approximate 10 μC/cm² for the ceramics samples at room temperature. It seems that there is a slight increase in the saturation polarization as increasing the Ca-concentration initially. This result is in agreement with the variation of Curie point with Ca-concentration. It also seems that the coercive field becomes larger with the increase in Ca-concentration.

On the other hand, the great enhancement of strain responses under an electric field is clearly observed for Ca-substituted ceramics as shown in Fig. 16 (b) & (c). For examples, the electric-field-induced strain for $x$=0 has a value of 0.054 % at $E$=10 kV/cm, for the same unipolar field, while it reaches a large value of 0.086% for $x$=0.10, which corresponds to an effective piezoelectric response of 860 pm/V. Figure 16 (d) demonstrates the variation of this effective piezoelectric response around room temperature with Ca-composition. It is clear that the large effective piezoelectric response with values higher than 800 pm/V has been observed in a wide composition range from $x$=0.06 to $x$=0.16 in the BCTZO ceramics, which is greatly larger than that obtained in the commercial PZT ceramics. Such extremely large electric-field-induced strain may be of great interest for the development of green piezoelectric ceramics. As shown in Fig. 16(d), the large electromechanical coupling effects occur around the *R-O* and *O-T* phase boundaries. This indicates that the polymorphic phase transitions play a critical role in the large piezoelectric response in the BCTZO solid solution.

## 4. Summary

Ca off-centering was demonstrated to play a critical role in stabilizing the ferroelectric phase and tuning the polarization states in the $(Ba_{1-x}Ca_x)(Ti_{1-y}Zr_y)O_3$ system. Two typical cases ($(Ba_{1-x}Ca_x)TiO_3$ with $y$=0 and $(Ba_{1-x}Ca_x)(Ti_{0.9}Zr_{0.1})O_3$ with $y$=0.1) have been studied. In both cases, atomic displacement due to Ca off-centering in the bulky Ba-site in the $ABO_3$ perovskite structure provides an approach to compensate the reduction of ferroelectricity due to chemical pressure, leading to keeping the Curie point nearly constant in the $(Ba_{1-x}Ca_x)TiO_3$ system and increasing the Curie point in the $(Ba_{1-x}Ca_x)(Ti_{0.9}Zr_{0.1})O_3$ system. The Ca off-centering effects are commonly observed to lead to the shift of the *R–O* and *O–T* phase transitions toward lower temperatures and the ferroelectric stability of the *T*-phase, resulting in the occurrence of quantum phase transitions with interesting physics phenomena at low temperatures in the $(Ba_{1-x}Ca_x)TiO_3$ system and the great enhancement of electromechanical coupling effects around room temperature in the $(Ba_{1-x}Ca_x)(Ti_{0.9}Zr_{0.1})O_3$ system over a large composition range.

## 5. Acknowledgment

We thank Prof. Shin-ya Koshihara & Dr. T. Shimizu in Tokyo Institute of Technology, Mr. T. Kosugi & Prof. S. Tsuneyuki in University of Tokyo, and Mr. Y. Kamai in Shizuoka University for the collaboration in this work. We also thank the support from Collaborative Research Project of Materials and Laboratory, Tokyo Institute of Technology.

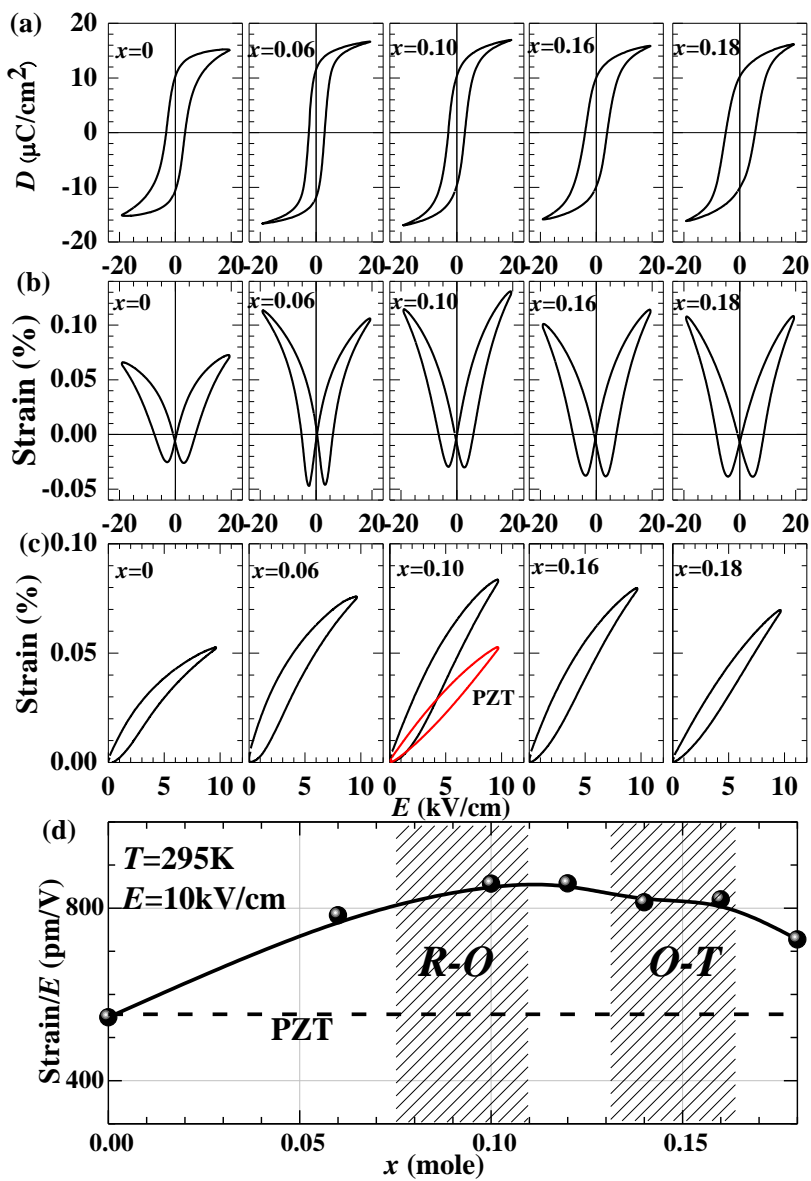

**Figure 16.** (a) Polarization, (b) bipolar-field and (c) unipolar-field strain under electric field measured at $T$=295 K for the $(Ba_{1-x}Ca_x)(Ti_{0.9}Zr_{0.1})O_3$ ceramics. For comparison, the response of commercially used PZT is also shown in (c). (d) Change of strain at $E$=10kV/cm with composition.

[Double click to insert bibliographysource here]